# *In silico* estimates of the free energy rates in growing tumor spheroids[*]

H Narayanan[1], S N Verner[2], K.L. Mills[3], R. Kemkemer[3] and K. Garikipati[2,4,5]

[1]Simula Research Laboratory, Oslo, Norway
[2]Department of Mechanical Engineering, University of Michigan, Ann Arbor, USA
[3]Department of New Materials and Biosystems, Max Planck Institute for Metals Research, Stuttgart, Germany
[4]Michigan Center for Theoretical Physics, and Center for Computational Medicine and Bioinformatics, University of Michigan, Ann Arbor, USA

**Abstract.** The physics of solid tumor growth can be considered at three distinct size scales: *the tumor scale, the cell-extracellular matrix (ECM) scale and the sub-cellular scale*. In this paper we consider the tumor scale in the interest of eventually developing a system-level understanding of the progression of cancer. At this scale, cell populations and chemical species are best treated as concentration fields that vary with time and space. The cells have chemo-mechanical interactions with each other and with the ECM, consume glucose and oxygen that are transported through the tumor, and create chemical byproducts. We present a continuum mathematical model for the biochemical dynamics and mechanics that govern tumor growth. The biochemical dynamics and mechanics also engender free energy changes that serve as universal measures for comparison of these processes. Within our mathematical framework we therefore consider the free energy inequality, which arises from the first and second laws of thermodynamics. With the model we compute preliminary estimates of the free energy rates of a growing tumor in its pre-vascular stage by using currently available data from single cells and multicellular tumor spheroids.

## 1. Background

The progression of a tumor involves (a) cell proliferation, (b) cell motility, (c) metabolism by which the cells consume glucose and oxygen and create byproducts, (d) mechanical interactions between cancer cells, the ECM and surrounding tissues, and (e) mass transport of chemical species to and through the tumor. Each of these processes has a physically-distinct contribution to the free energy rate in the developing tumor. Complex biophysical interactions between these processes are more broadly observable at the tumor scale than in single-cell studies. Additionally, as we demonstrate in this communication, tumor scale studies have the potential of identifying the relevant questions regarding energy rates that must be considered at the lower, cell-ECM and sub-cellular scales. Using the tumor scale studies, it is of interest to track the free energy rates and thereby gain a system-level understanding of the processes listed above in developing tumors. More broadly, we argue in the Discussion of this paper that there is an interest in combining free energy studies at the tumor, cell-ECM and sub-cellular scales. This will reveal how the energetics change with time and state of the tumor between these different scales and between the processes underlying tumor growth to affect the progression of the cancer.

*1.1 Biochemical dynamics and mechanics of tumors; a role for free energy*
The progression of cancer can be framed in terms of pathways of energy consumption. This idea has been applied at the molecular scale to the identification of specific metabolites in the cancer cells, which implicates certain pathways of cellular metabolism (Denkert et al., 2008). Whether these pathways are normal or altered, the cell directs the associated free energy rates to some of its functions: proliferation,

---

motility and mechanical interactions. Conversely, the cell's chemical stores of energy (in the form of ATP) are replenished by glucose metabolism in the presence of oxygen—a process involving mass transport over the extent of the tumor and reactions within its cells. The physics of solid tumors at the cell-ECM scale thus comprises a broad range of cellular functions that are not apparent at the sub-cellular (e.g., organelle) scale. Additionally, the progression of cancers that produce solid tumors is not determined solely by the bio-chemo-mechanics within single cells, or the interactions of a few cells with each other and with the ECM. While these local effects are very important, they depend also on spatio-temporal conditions and cooperative effects that are particularly apparent at the tumor scale. This is one reason for the interest in tumor-scale studies of the physics of cancer (Heiden et al., 2009; Kumar & Weaver, 2009; Levental et al.,2009; Paszek et al., 2005; Gatenby & Gillies, 2004; Gordon et al., 2003; Guiot et al., 2003; Koike et al., 2002; Bell et al., 2001; Freyer, 1998; Helmlinger et al., 1997; Weaver et al., 1997; Casciari et al., 1992; Bourrat-Floecke et al., 1991). At this scale, cells are represented by concentration fields. The concentrations increase and decrease (by cell proliferation and death, respectively), the cells undergo transport (cell motility), they deposit and degrade the ECM and develop mechanical stress (from cell-cell and cell-ECM mechanical interactions). The cells also consume glucose and oxygen as well as create by-products, all of which also are represented as concentrations at the tumor scale.

There are two main features of tumor-scale physics that have been studied for their influence on the progression of the cancer: (a) The *biochemical dynamics*, by which we mean the changing concentration fields of cells, ECM, oxygen, glucose and by-products, which result from the processes of cell proliferation and death, cell motility and metabolism (Heiden et al., 2009; Gatenby & Gillies, 2004; Bell et al., 2001; Freyer, 1998; Groebbe & Mueller-Klieser, 1996; Casciari et al., 1992; Freyer et al., 1991; Bourrat-Floecke et al., 1991; Freyer & Sutherland, 1986, 1985; Franko & Sutherland, 1979; Weinhouse, 1956; Warburg et al., 1927). (b) The *mechanics* of interactions among cancer cells and between cells and the ECM. The forces involved in these mechanical interactions have been implicated in cell proliferation, motility and signaling (Kumar & Weaver, 2009; Butcher et al., 2009; Chang et al., 2008; Suresh, 2007; Kaufman et al., 2005; Padera et al., 2004; Gordon et al., 2003; Koike et al., 2002; Helmlinger et al., 1997). All physical processes at the sub-cellular and cell-ECM scales manifest themselves in either the biochemical dynamics (as defined above) or mechanics at the tumor scale, and the operation of all these processes involves free energy changes. Therefore, free energy change at the tumor scale is a universal measure for quantification and comparison of the physical processes that govern the cancer's progression and perhaps is the only measure that unifies the biochemistry and mechanics of tumor growth.

As the above references suggest, the biochemical dynamics and mechanics of tumors have mostly been studied in isolation with a focus on the biochemical dynamics that has only recently begun to yield some ground to studies of the mechanics of tumors. There have been only a few studies—all focused on mathematical modeling—where these two aspects have been studied in combination (Cristini et al., 2009; Frieboes et al., 2006; Zheng et al., 2005; Drasdo & Höhme, 2005; Jackson & Byrne, 2002). None, however, has considered the free energy rates associated with the biochemical dynamics and mechanics of tumors.

*1.2 Studies of the biochemical dynamics of tumor growth*
Experimental studies of the biochemical dynamics of cancer have maintained a focus on cell proliferation rates under varying concentrations of glucose, oxygen and $H^+$ ions (Gatenby & Gillies, 2004; Freyer, 1998; Groebbe & Mueller-Klieser, 1996; Casciari et al., 1992; Freyer et al., 1991; Groebbe & Mueller-Klieser, 1991; Tannock & Kopelyan, 1986; Mueller-Klieser et al., 1986; Sutherland et al., 1986; Freyer & Sutherland, 1986, 1985; Franko & Sutherland, 1979; Warburg et al., 1927) and in some instances, of lactate (Bourrat-Floecke et al., 1991). While there have been a few *in vivo* studies among them, the majority of these studies have used multicellular tumor spheroids—an *in vitro* cancer model corresponding to initial pre-vascular or inter-vascular microregions of *in vivo* tumors. Tumor spheroids derived from chosen cancer cell lines have been grown from sizes of ~50 μm in growth medium perfused with glucose and oxygen. After an initial exponential growth phase (so-called because the cell count increases exponentially in time), the size reaches a plateau resulting in a characteristic sigmoidal shape of the size vs. time curve, which has been represented by an empirical fit termed the Gompertzian equation (Gompertz, 1825).

The sigmoidal shape of growth curves has been of interest to theoretical biologists. On the basis of hydrodynamic scaling laws and fractal branching of the terminal vasculature, West et al. (2001, 2002) proposed that the total metabolic rate of an organism is proportional to the $3/4^{th}$ power of its mass, and derived growth laws that can match the time progression of the size of biological systems across many orders of magnitude. Guiot et al. (2003) demonstrated that these growth laws are able to represent the sigmoidal growth curves of tumors. While it is an elegant approach that is also energy-based, this scaling law applies to the tumor as a whole. It therefore does not shed light on the detailed distribution of free energy rates between the many tumor-scale processes that we have described in Section 1.1 and that govern the physics of cancer. Focusing on these physical processes, however, it has been shown (see the references at the beginning of the previous paragraph) that the plateau in growth is reached because, as the tumor spheroid grows, the biochemical dynamics become diffusion-limited: glucose and oxygen do not diffuse rapidly enough to the center to supply the cells as they continue to deplete these chemicals by metabolic activity. The cells slip into a quiescent state and eventually die, forming a necrotic core, typically when the tumor spheroid has attained a diameter of ~500 μm. This is the end of the exponential growth phase. The tumor spheroid's size grows at a vanishingly small rate as the net cell proliferation rate tends to zero. For spheroids derived from different cell lines, but growing in the same environment (in terms of availability of nutrients and stiffness of surrounding media), this limiting size is determined by the cell type (Folkman & Greenspan, 1975). The next stage of tumor progression *in vivo* is vascularization, which gives the tumor a new lease on life as newly-formed blood vessels begin to supply glucose and oxygen to the cells.

*1.3 The lately-emerging mechanics of tumors*
Evidence that mechanics also affects the tumor's progression has appeared more recently. Helmlinger et al. (1997) found a suppression of growth of tumor spheroids derived from colon cancer LS174T cells when subjected to compressive stress by an encapsulating hydrogel. In a follow-up study Koike et al. (2002) demonstrated that externally-applied mechanical stress aids the formation of multicellular tumor spheroids in the highly metastatic Dunning R3327 rat prostate carcinoma AT3.1 cells, while the less metastatic AT1 cells formed spheroids even without the applied stress.

Chang et al. (2008) showed that in four different cell lines shear stress led to cell cycle arrest in the G2/M phase. This result was associated with increased expression of cyclins B1 and $p21^{CIP1}$, and decreased expression of cyclins A, D1 and E, cyclin-dependent kinases (cdk) -1, -2, -4, -6 and $p27^{KIP1}$, as well as decreased cdk-1 activity. Reviews by Kumar & Weaver (2009), Butcher et al. (2009) and Suresh (2007) have pointed to the decreased stiffness and altered cytoskeletal rheology of cancer cells from several different cell lines when compared with normal cells. These phenotypes promote greater motility and therefore probably favor metastasis of the cancer.

Cells also impose traction forces on the ECM, and Gordon et al. (2003) found that larger traction forces near the edge of tumor spheroids with the human U87MGmEGFR glioblastoma cell line led to greater depths of invasion into the ECM. Many cell types form focal adhesions with the ECM, and the force developed in the actin cytoskeleton is regulated by a dynamic interaction between focal adhesions, the cytoskeleton and the ECM (Geiger & Bershadsky, 2001, and references therein). It is hypothesized that the force so-developed regulates the expression and activity of many proteins by mechanisms that are yet undiscovered, and that this chemo-mechanical regulation may influence the chemical signaling in cancer cells (Kumar & Weaver, 2009, and references therein). Butcher et al. (2009) have also pointed to altered

"mechanoreciprocity" (the development of force within the cell in response to ECM-imposed strain) by which higher-than-normal forces are applied to cell-cell junctions causing them to lose their integrity, thereby aiding in tissue invasion. Weaver et al. (1997) found that mechanical interactions between integrins and the ECM altered the phenotype of human breast cancer cells and that under certain interventions these cells reverted to the normal phenotype. Padera et al. (2004) demonstrated that mechanical stress created by growing tumors compresses blood vessels supplying the tumors and thereby interferes with the delivery of both nutrients and drugs.

The physical processes that manifest themselves in the biochemical dynamics and mechanics of tumors are of interest because they have all been found to influence the progression of the cancer. Our aim is to study them in the context of the free energy changes that are caused by these processes. Toward this goal, Section 2 provides an outline of our mathematical models of these processes at the tumor scale and the theoretical basis for considering the associated free energy rates. Section 3 is a brief discussion of the experiments that we have initiated to support the computational studies. Section 4 presents computational studies that serve as preliminary estimates of free energy rates in a growing, pre-vascular tumor. A discussion and conclusions are presented in Section 5.

**2. The continuum model for the physics of growing tumors**
Our mathematical formulation for the physics of tumor growth is drawn from a broader treatment that we have developed for the growth and remodeling of biological tissue. It is based on the continuum theory of mixtures and has been detailed in Garikipati et al. (2004, 2006) and Narayanan et al. (2009).

*2.1 The biochemical dynamics of tumor growth*
The biochemical dynamics of the tumor are governed by a system of coupled reaction-transport partial differential equations (PDEs) for the concentrations of cells $\rho^c$, ECM $\rho^e$, oxygen $\rho^o$ and glucose $\rho^g$.

$$\frac{\partial \rho^c}{\partial t} = \pi^c - \nabla \cdot (-D^c \nabla \rho^c) \tag{1}$$

$$\frac{\partial \rho^e}{\partial t} = \pi^e \tag{2}$$

$$\frac{\partial \rho^o}{\partial t} = \pi^o - \nabla \cdot (-D^o \nabla \rho^o) \tag{3}$$

$$\frac{\partial \rho^g}{\partial t} = \pi^g - \nabla \cdot (-D^g \nabla \rho^g) \tag{4}$$

Here, $D^\alpha$ is the diffusivity of the corresponding species, where $\alpha = c, o$ or $g$. The diffusive term in Equation (1) models the random motion of cells in the absence of a chemotactic or haptotactic driving force.[6] In Equation (2) the ECM does not undergo transport, while in Equations (3) and (4) oxygen and glucose, respectively, undergo diffusive transport. Source terms modeling the proliferation rate of cells, the ECM production rate, and the consumption rates of oxygen and glucose, are $\pi^c, \pi^e, -\pi^o$ and $-\pi^g$, respectively.

---

[6] Directed motion of the cells can arise under a chemotactic driving force that causes them to migrate away from a toxic by-product, in the direction of the vascular supply, or toward nutrients. Alternately, haptotaxis may influence them to migrate toward a more abundant ECM. These responses are, however not observed in the early, pre-vascular and pre-necrotic stages that we are considering in this work. Mathematically-speaking, chemotactic and haptotactic cell motion would be modeled by convection terms in Equation (1).

The cell proliferation rate (expressed as a mass rate per unit volume with units of mg-cc$^{-1}$-sec$^{-1}$) has been chosen to model the initial, exponential stage of tumor growth.

$$\pi^c = \frac{\rho_0^c}{J} \frac{\log 2}{t_D} 2^{t/t_D} = \frac{\rho_0^c}{J} \frac{1}{\tau} e^{t/\tau}, \tag{5}$$

where $\rho_0^c$ is the initial cell concentration, $t_D$ is the cell doubling time, $\tau = t_D/\log 2$, and $J$ is the ratio by which an infinitesimally-small volume of the tumor has deformed and grown. This last factor has been made more precise in the discussion of mechanics that follows in Section 2.2. The cell doubling time is dependent on the oxygen and glucose concentrations (Casciari et al., 1992), and the pH ($\rho^{H^+}$) of the medium (Casciari et al., 1992; Bourrat-Floecke, 1991). We have used the equation proposed by Casciari et al., (1992) for $t_D$ as a function of $\rho^o, \rho^g$ and $\rho^{H^+}$.

$$t_D = \frac{t_D^{opt}}{0.014}\left(\frac{\rho^o + 7.3 \times 10^3}{\rho^o}\right)\left(\frac{\rho^g + 1.8 \times 10^{-2}}{\rho^g}\right)\left(\rho^{H^+}\right)^{0.46}, \tag{6}$$

where $t_D^{opt}$ is the optimal doubling time found to be attained at $\rho^o = 1.79 \times 10^{-3}$ mg-cc$^{-1}$, $\rho^g = 0.99$ mg-cc$^{-1}$ and $\rho^{H^+} = 5.62 \times 10^{-8}$ mg-cc$^{-1}$ (pH = 7.25). Equation (6) was compared with data in Freyer & Sutherland (1985, 1986) in addition to those in Casciari et al. (1992). A least squares fit returned a value of $R^2 = 0.75$. While in Casciari et al. (1992) $t_D^{opt} = 11$ hours, we have left this factor as a parameter to match our preliminary tumor spheroid growth experiments (see Section 3.1). Note that, at fixed $\rho^{H^+}$, the time required for doubling the number of cells becomes unbounded as $\rho^o, \rho^g \to 0$, implying that the cell proliferation rate, $\pi^c \to 0$. The vanishing oxygen and/or glucose concentration provides a basis for modeling cell necrosis as we discuss later. While Casciari et al. (1992) used fixed values of $\rho^o, \rho^g$ over the tumor spheroid in their experiments, we have assumed that Equation (6) holds pointwise over the tumor spheroid. Importantly, as $\rho^o, \rho^g$ vary with time and space, the calculated cell doubling time also varies. The optimal doubling time, $t_D^{opt}$, however, is fixed for a chosen cell line.

The ECM production rate has been modeled to be proportional to the cell concentration:
$$\pi^e = A\rho^c, \tag{7}$$
with $A$ being a constant of proportionality.

The oxygen and glucose consumption rates also were adapted from Casciari et al. (1992) to be consistent with the different units used here, and scaled by the local cell concentration to be expressed as mass rates per unit volume. The resulting rate functions take on field values that vary over time and space, and have the forms:

$$\pi^o = -\rho^c \left(7.68 \times 10^{-7} + \frac{3.84 \times 10^{-15}}{\rho^g (\rho^{H^+})^{0.92}}\right)\left(\frac{\rho^o}{\rho^o + 1.47 \times 10^{-4}}\right), \tag{8}$$

$$\pi^g = -\rho^c \left(1.14 \times 10^{-10} + \frac{3.65 \times 10^{-17}}{\rho^o}\right)\left(\frac{\rho^g}{\rho^g + 7.21 \times 10^{-3}}\right)\left(\frac{1}{\rho^{H^+}}\right)^{1.2}, \tag{9}$$

with units of mg-cc$^{-1}$-sec$^{-1}$. The dependence of $\pi^o$ on $\rho^o$, and of $\pi^g$ on $\rho^g$ is "Michaelis-Menten-like", giving rates that vary monotonically from zero to a maximum asymptotic value as the respective concentrations increase from $\rho^o, \rho^g = 0$. Also note that $\pi^o$ is inversely proportional to $\rho^g$, and $\pi^g$ is

*In silico* estimates of free energy changes in growing tumor spheroids inversely proportional to $\rho^o$. These trends are reflected in the data of Freyer & Sutherland (1985, 1986) and Casciari et al. (1992).

*2.2 Mechanics of the tumor*
The PDEs for reaction-transport of cells and ECM are coupled with the quasi-static balance of momentum that governs the mechanics of the tumor. For the purpose of mechanics the tumor is treated as a soft material consisting of cells and ECM. The PDE for quasi-static balance of momentum is:
$$\nabla \cdot \sigma = 0, \tag{10}$$
where $\sigma$ is the Cauchy stress, having a passive viscoelastic contribution from the mechanical response of the ECM and cells, and an active contribution due to cell traction. The total stress is therefore written as, $\sigma = \sigma_{el} + \sigma_v + \sigma_a$ the subscripts denoting "elastic", "viscous" and "active", respectively.

The deformation gradient tensor is $F = I + \partial u/\partial X$, where $I$ is the second-order identity tensor, $u$ is the displacement vector and $X$ is the reference position. The ratio of current (deformed and grown) to reference (undeformed and initial) volume is $J = \det F$. Since deformation results from both elastic strain and growth, we can write $F = F_{el}F_{gr}$, where $F_{gr}$ is an isotropic tensor and represents the kinematic growth caused by cell proliferation and ECM deposition. Accordingly we write
$$F_{gr} = \left((\rho^c + \rho^e)J/(\rho_0^c + \rho_0^e)\right)^{1/3} I, \tag{11}$$
where $\rho_0^c$ is the initial cell concentration and $\rho_0^e$ is the initial ECM concentration. We also define the volume change ratio due to elastic strain, $J_{el} = \det F_{el}$

The elastic part of the stress is obtained from the standard relation for a hyperelastic material $\sigma_{el} = J_{el}^{-1} F_{el}(\partial W/\partial C) F_{el}^T$, where $W$ is the Mooney-Rivlin strain energy function and $C$ is the elastic right Cauchy-Green tensor at a given material point. These quantities are related to the deformation of the solid tumor as $W = W(C)$, where $C = F_{el}^T F_{el}$. Specifically, the Mooney-Rivlin model is
$$W(C) = \frac{1}{2}\kappa(J_{el} - 1)^2 + \frac{1}{2}\mu(\bar{I}_1 - 3) \tag{12}$$
Here $\kappa$ and $\mu$ and are, respectively, the bulk modulus and shear modulus, the latter in the limit of infinitesimal strain, and $\bar{I}_1$ is the first principal invariant of $\bar{C} = J_{el}^{-2/3} C$ defined as $\bar{I}_1 = trace(\bar{C})$. The viscous stress is written as
$$\sigma_v = (J_{el})^{-5/3} F_{el} Q F_{el}^T, \tag{13}$$
where $Q$ is a stress-like quantity that is governed by the ordinary differential equation
$$\dot{Q} + \frac{Q}{\tau} = \frac{\gamma}{\tau} dev\left[2\frac{\partial \bar{W}}{\partial \bar{C}}\right], \tag{14}$$
with $\tau$ being an intrinsic relaxation time, and $\bar{W} = \frac{1}{2}\mu(\bar{I}_1 - 3)$. For the tumor growth phenomena that will be studied with this model, the strain rate is set by the rate of volume growth, which happens over a time scale of 1—30 days. This rate therefore is of the order of $10^{-7}$ sec$^{-1}$. In contrast, typical relaxation times of soft tissue are in the range of 1000 sec, giving a larger intrinsic rate of $1/\tau \sim 10^{-3}$ sec$^{-1}$. The viscous effects are therefore negligible, and the viscous stress, $\sigma_v$, has been set to zero in this model. Similarly, we have also found that the intrinsic rates of the gels that encapsulate tumor spheroids in our preliminary experiments are large compared with the strain rates due to growth, lending further support to the neglect of viscous effects in tumor growth phenomena (see Section 3.2).

The active stress arises due to the tensile traction imposed by the cells on the ECM, $\sigma_a = \beta(\rho^c/\rho^c_{max})^{2/3} I$, which is isotropic, as indicated by the second-order identity tensor, *I*, with $\beta$ being a measure of the maximum traction developed. The exponent of 2/3 converts the volume concentration ratio to an area concentration ratio. According to this model an isotropic stress $\beta$ is developed at $\rho^c = \rho^c_{max}$. The value of $\beta$ has been determined from the stress of 5.5 kPa measured on focal adhesions (Balaban et al., 2001). This value was scaled up geometrically by accounting for the typical numbers and sizes of focal adhesions in cells to obtain our estimate for $\beta$. Table 1 lists additional parameters used in the model, with citations or remarks on how the corresponding parameter was obtained.

*2.3 Free energy rates in the growing tumor*
Our mathematical model extends to the thermodynamics, which encompasses effects arising from the biochemical dynamics and mechanics of the tumor. The application of the first and second laws of thermodynamics to our mathematical model leads to an inequality that governs the free energy rates of the growing tumor. The derivation has been detailed in Garikipati et al. (2004, 2006) for tissue undergoing growth and remodeling.:

$$\rho^c \dot{\psi}^c_{chem} + \pi^c \psi^c + \pi^e \psi^e - \sigma : d_{gr} + \rho^c \nabla \psi^c \cdot v^c + \pi^g \psi^g \leq 0 \tag{15}$$

Here, $\psi^\alpha$ is the free energy per unit mass of the constituent $\alpha$ (=c, e, o, g), $\psi^c_{chem}$ is the chemical free energy of the cells,[7] $d_{gr} = F_{el}\dot{F}_{gr}F_{gr}^{-1}F_{el}^{-1}$ is the rate of deformation tensor due to growth, and $\rho^c v^c$ is the flux of motile cells given by $\rho^c v^c = -D^c \nabla \rho^c$. The mass-specific chemical free energy of cells changes at a rate given by

$$\dot{\psi}^c_{chem} = -B_{cell} m_{cell}^{-1} + \frac{\pi^g \psi^g}{\rho^c} \tag{16}$$

where the $B_{cell}$ is a constant metabolic power output for all mammalian cells in culture as shown in West et al. (2002) and $m_{cell}$ is mass of a single cell (see Table 1 for values). The second term in (16) is exactly the rate at which chemical energy is extracted by consuming glucose. By adopting this form we have assumed that the energy gained from glucose is stored in the cells without losses.

The first term on the left hand-side of the free energy inequality represents the rate at which chemical free energy density is changing due to usage and storage within the cells. The remaining terms are, respectively, the free energy density rates due to cell proliferation, ECM production, tumor growth against stress, cell motion, and due to consumption of glucose. The inequality requires that the sum of these rate terms be negative, meaning that the total free energy density is decreasing. Each of the terms in the free energy inequality can be computed from our model using experimentally-determined parameters that represent a specific tumor cell line, the environment of the tumor cells (such as whether they form a tumor spheroid surrounded by a gel, as in Section 4), and initial and boundary conditions. The evaluation of these terms provides a quantitative comparison of the free energy rates that occur in these distinct tumor-scale processes. As mentioned in Section 1, we emphasize that the biochemical dynamics and mechanics of the tumor are subjected to unified treatment in this comparison of free energies.

It also is useful to rewrite (15) as

---

[7] For the physical processes considered here, the total free energy is the sum of the chemical free energy and the mechanical strain energy.

*In silico* estimates of free energy changes in growing tumor spheroids

$$\rho^c \dot{\psi}^c_{chem} + \pi^c \psi^c + \pi^e \psi^e \leq \sigma : d_{gr} - \rho^c \nabla \psi^c \cdot v^c - \pi^g \psi^g \quad (17)$$

If the mechanical power represented by the first term on the right hand-side is positive in sign, the corresponding free energy change is not stored, and therefore represents a mechanism of dissipation. If the change in free energy represented by the rate $-\rho^c \nabla \psi^c \cdot v^c$ is negative, it also represents a dissipative mechanism. A positive value of $-\pi^g \psi^g$ represents the energy extracted by consuming glucose. The left hand-side represents mechanisms of energy storage. In this form, therefore, the inequality specifies that the rate of free energy storage is less than the rate of energy loss due to dissipation and glucose consumption.

*2.4 Numerical implementation of the mathematical model*
The five coupled PDEs (1—4) and (10) are solved by the finite element method. Several mathematical complexities can arise in the solution of this coupled system of PDEs, and we have discussed their treatment at length in Narayanan et al. (2009). Briefly, the reaction-transport PDEs are integrated in time by the midpoint rule. A mixed finite element method employing the displacement-pressure formulation has been used for the mechanics of the soft, nearly-incompressible tumor. Of special relevance to this study has been the need to rapidly test formulations with different forms of the PDEs, response functions and constitutive models, and for a range of initial and boundary conditions. For this purpose we have adopted the multiphysics modeling code, Comsol.[8] A typical computation with ~5000 finite elements run to 20 days of physical time took 2 hours of wall time to run on an IBM Thinkpad T43 laptop with 2 GB RAM and a processor speed of 2 GHz..

**3. Experimental methods**
Experimental work with cancer cells and the tumor spheroids that they form will provide the values for the essential parameters of our model. While data are available for certain aspects of chosen cell lines and tumor spheroid systems, there is a need for comprehensive data that are consistent in the sense that they are obtained for the same cell line(s) and tumor spheroid system(s). Preliminary experimental work has included the optimization of tumor spheroid production, the seeding of the resulting tumor cells in hydrogels, and the observation of the subsequent growth of the tumors over time periods up to one month. The tumor spheroids have been grown in hydrogels of various concentrations and we have probed the gels' mechanical properties.

*3.1 Cancer cell culture maintenance, tumor spheroid production, and growth*
For preliminary experiments, we have chosen to work with three different adherent epithelial cancer cell lines: human colon adenocarcinoma (LS174T), human cervix carcinoma (HeLa), and human breast adenocarcinoma (MCF-7). Cell subculturing, prior to tumor spheroid production is carried out in tissue-culture flasks with a cell attachment-treated surface area of 75 cm$^2$. The different culture conditions used for each cell line are as follows. For LS174T cells: BioWhittaker EMEM (Lonza) containing L-glutamine plus an additional 10% fetal bovine serum (FBS), 1% NEAA, 1% penicillin and 1% streptomycin (1% P/S). For HeLa cells: GIBCO RPMI 1640 (Invitrogen) containing L-glutamine plus an additional 10% FBS and 1% P/S. For MCF-7 cells: GIBCO DMEM (Invitrogen) containing 4.5 g/L glucose plus an additional 10% FBS and 1% P/S. Cells are detached from the tissue culture flask surface prior to growing to confluence using GIBCO trypsin EDTA (0.05%, Invitrogen) and either split or transferred to an experimental platform. Cells being maintained in culture conditions or during experiments are stored in incubators with a controlled environment of 37°C, high humidity, and 5% CO$_2$.

In all experiments, tumor spheroid formation is initiated using the hanging drop method. Cells are suspended in their culture medium at a concentration of 5000 cells-cc$^{-1}$. Drops of 6.5 μl, containing the cell suspension, are placed with a pipette on the underside of the cover of a Petri dish and the cover is then inverted and replaced on the dish. Gravity and the surface tension of the liquid confine the approximately 10-50 cells per drop to a small quasi-spherical volume where the formation of cell-to-cell attachments is encouraged. The time necessary for the formation of an agglomerate of cells, which adhere to each other and produce their own ECM, within the drops is cell-line dependent. For example, LS174T cells form spheroids within approximately 12 hours, whereas MCF-7 cells require 2-3 days.

---
[8] www.comsol.com

Spheroids are transferred into agarose hydrogels using a gel overlay method. Agarose hydrogels, in varying concentrations of agarose, provide different levels of 3D mechanical support for the tumor spheroids without biochemical interactions between the tumor cells and their environment. Whereas tumor-environment interactions are important, in the initial stages of our study we seek to restrict the energy flow to only mechanical interactions with the environment and processes internal to the tumor spheroid. By doing this, we can concentrate on identifying the effects of nutrition (i.e., oxygen and glucose) and stiffness of the environment without the confounding effects of an additional set of ECM molecules. In later phases of this research, adhesive and interactive 3D environments will provide additional insights on the very important biochemo-mechanical feedback that a tumor has with its surrounding ECM. For preparation of the hydrogels, the appropriate amount of agarose powder is mixed with deionized water and heated in a microwave oven to fully dissolve the powder and create a stock solution of 2.0 wt.% agarose hydrogel. When the agarose stock solution has cooled to at least 37°C it is mixed with cell-culture medium in order to obtain the desired final agarose concentration. A thin layer (approximately 2 mm thick) of un-gelled agarose and cell-culture medium solution is then placed in the bottom of a cell culture well. Shortly before the agarose forms into a gel, the spheroids are individually transferred just under the surface of the gel using a 10 μl pipette tip. To ensure complete coverage of the spheroid, an additional layer of un-gelled agarose solution is placed on top of the first. The agarose is allowed to gel at room temperature for 20 minutes after which cell culture medium is added to each well.

After the spheroids have been seeded in the gels, we have either continuously monitored the development of single tumor spheroids for time periods up to one week or, for more long-term statistical data, acquired images of many tumor spheroids (approximately 20 per experiment) at 48-hour intervals for up to one month. Continuous monitoring is performed on an inverted microscope (Zeiss Axiovert 200) using phase contrast and DIC techniques with images being captured every 15 minutes. An incubator housing is placed over the stage of the microscope to maintain a humid environment at 37°C and 5% $CO_2$.

Preliminary experiments have shown that the LS174T cell line agglomerates readily and rapidly to form spheroids under the growth conditions described above. In a typical, successful experiment where the development of the tumor was monitored continuously, an LS174T tumor grew from a radius of ~50 μm to ~200 μm over 7 days, a quadrupling of the radius. Assuming uniform cell concentration in the spheroid this corresponds to an increase in number of cells by a factor of 64 ($2^6$), which corresponds to 6 doublings of the cell population in 7 days or a doubling time $t_D$ ~1.16 days. Figure 1 shows a sequence from one of our preliminary tumor spheroid growth experiments. The growth conditions for oxygen and glucose corresponded to initial and boundary values $\rho^o = 7.336 \times 10^{-3}$ mg-cc$^{-1}$ (equilibrated with partial pressure corresponding to 20% oxygen) and $\rho^g = 0.99$ mg-cc$^{-1}$. This value of $\rho^o$ is $5\times$ the optimal value found by Casciari et al. (1992) and the value of $\rho^g$ is optimal. For our preliminary studies we have assumed that the doubling time is also optimal with respect to oxygen. On this basis we have used an optimal cell doubling time of $t_D^{opt} = 1.16$ days.

*3.2 Hydrogel characterization*
As previously mentioned, the medium with which we embed agglomerates of cancer cells for the growth and characterization of tumor spheroids is agarose hydrogel (Type VII, Low gelling temperature, Sigma

*In silico* estimates of free energy changes in growing tumor spheroids

Chemical Co., St. Louis, MO, USA), in concentrations of 0.5 – 2% (wt·vol$^{-1}$) agarose. The mechanical properties of agarose hydrogels are sensitive to supplier, type, and preparation method (Luo & Shoichet, 2004; Stolz, 2004). Additionally, initial experiments show that the behavior of agarose is rate dependent, likely due to viscoelastic contributions of the polymer and poroelastic effects of the hydrating fluid. Characterization of the macroscopic mechanical behavior of hydrated agarose gels is being carried out with unconfined compression and stress relaxation tests using an MTS NanoBionix Test System.

Unconfined compression and stress relaxation are performed between aluminum platens: one platen is fitted with a Plexiglas cylinder containing the agarose sample submerged in a bath of hydrating fluid. A brief description of sample preparation for mechanical testing of agarose gels is as follows. After gelling is complete, cylinders are stamped out of approximately 2 mm-thick sheets using the large-diameter end of a Pasteur pipette. The resulting nominal diameter of the agarose cylinders is 5 mm. The top platen compresses the gel with strain rates between $1 \times 10^{-4}$ sec$^{-1}$ and $1 \times 10^{-3}$ sec$^{-1}$ to maximum strains between 5 and 20% at which point the crosshead displacement is halted and the stress is continually measured for an additional two minutes before unloading at the same rate.

From our preliminary mechanical characterization of the gels we have obtained apparent moduli of the order of 1—25 kPa with a recognizable strain rate effect (see Table 2). We also found relaxation times of ~200 sec in the stress relaxation tests on gels. This corresponds to intrinsic rates of ~0.005 sec$^{-1}$ which are many orders of magnitude greater than the growth-induced volume strain rate of ~$10^{-7}$ sec$^{-1}$. For this reason we have neglected the viscous effect in gels for our computations as indicated in Section 2.2, and guided by our preliminary experimental results presented in Table 2, we have estimated a value of 6 kPa for the Young's modulus of 2% agarose gels for small strains. Since the gels have a high water content (98-99.5%), they are found to be nearly incompressible, which we have modeled by taking the Poisson ratio to be 0.49 in the small strain regime. This gives a bulk modulus $\kappa = 100$ kPa and shear modulus $\mu = 2013$ Pa (see Table 1).

## 4. Numerical estimates of free energy rates

Figure 2 introduces the reader to the main fields that are solved for in the formulation of initial and boundary value problems (IBVPs) of tumor physics. The computation is of a growing tumor spheroid, shown here after 20 days of growth, encapsulated by a gel. The parameters used in this computation appear in Table 1. The extent of the tumor spheroid is revealed by the central, high concentration of cells. The initial distribution of $\rho^c$ is uniform at 510 mg-cc$^{-1}$ (equivalent to 170 cells in a volume $100 \times 100 \times 100 \mu$m$^{-3}$) over the tumor spheroid. The higher values of $\rho^c$ after 20 days of growth imply a higher cell packing density by a factor of $\sim 2.6$, corresponding well with the observations of Helmlinger et al. (1997). Outside of this high concentration lies the encapsulating gel where $\rho^c$ decreases sharply to close to zero. This introductory computation shows that the spheroid has grown to a radius of ~120 μm (demarcated by the central high values of $\rho^c$) over 20 days ($1.728 \times 10^6$ sec) starting from a radius of 50 μm as the cells produce ECM and proliferate to fill the newly laid down matrix.[9] As a result the gel has been deformed. The cells also consume oxygen and glucose. The initial distributions of oxygen and glucose were uniform over the tumor spheroid and gel: $\rho^o = 7.42 \times 10^{-4}$ mg-cc$^{-1}$, which is ~10% of the value in our preliminary experiments. The oxygen concentration in our experiments was equilibrated with the partial pressure of 20% oxygen in gas, which corresponds with the partial pressure of oxygen in the atmosphere. The lower $\rho^o$ was chosen for the computations because oxygen concentrations in tissues are usually found to be at most 20% of the value that equilibrates with the partial pressure of oxygen in the atmosphere. The initial glucose concentration was $\rho^g = 0.99$ mg-cc$^{-1}$, the same value as in our preliminary experiments. These concentration values were maintained as boundary conditions at the outer surface of the gel. Many studies were carried out, of which we have presented only the most relevant results here and in the parametric studies that follow.

---

[9] In the computations the tumor spheroid has grown to a lesser extent than in our exemplary preliminary experiment with LS174T cells. More experiments are needed, however, to obtain statistical error bounds on growth rates. Also note that the oxygen concentration in the computations was ~10% of the experiments, which leads to a lower growth rate according to Equation (6). A 2% agarose gel was modeled in the computations instead of the 0.5% agarose gel in the tumor spheroid growth experiments. The stiffer gel in the computations also produces a greater elastic constraint on growth of the tumor spheroid.

The computation proceeds with diffusion and consumption of oxygen and glucose, cell proliferation and ECM production, and the resultant growth of the spheroid. Oxygen and glucose are depleted by the cells as shown by the upper surface plot of $\rho^o$ and $\rho^g$. By 20 days $\rho^o$ and $\rho^g$ remain high only in an outer rim of cells, and decrease toward the center. As discussed in Section 1, the depletion of oxygen and glucose has been found to influence the onset of necrosis in experiments on tumor spheroids (Gatenby & Gillies, 2004; Bell et al., 2001; Freyer, 1998; Groebbe & Mueller-Klieser, 1996; Casciari et al., 1992; Freyer et al., 1991; Tannock & Kopelyan, 1986; Mueller-Klieser et al., 1986; Sutherland et al., 1986; Freyer & Sutherland, 1986, 1985; Franko & Sutherland, 1979; Warburg et al., 1927). Later in the paper we show a more pronounced depletion of oxygen and glucose in a computation that models more aggressively proliferating cells at a later time in the spheroid's growth.

Figure 3 shows the sum of the rate quantities that make up the left hand-side of the free energy inequality (15) for the case modeled in Figure 2. Recall that each term is the rate of change of free energy density from a specific mechanism as discussed in Section 2.3. This sum of the rates is negative at all points over the tumor spheroid and gel, indicating a decrease of free energy density in accordance with the free energy inequality. Note the order of magnitude difference in the rate of change of the free energy density between the tumor spheroid and gel. This difference is related to the cell distribution, which is high in the tumor spheroid but drops sharply to zero in the gel. Consequently, all cellular physical processes are significant only in the tumor spheroid, and comparatively negligible in the gel.[10] Diffusive transport of oxygen and glucose, and mechanical deformation are the only physical processes of significance in the gel. Due to the uniformity of $\rho^o$ and $\rho^g$ over the gel, the mass-specific free energy densities of oxygen and glucose are also uniform, and the change in free energy density due to their diffusion can be neglected. Since $\rho^c$ drops sharply to a very low value in the gel, the growth rate is negligible there, and the change in free energy density associated with growth against stress is also negligible over the gel.

The surface plots in Figures 4a—4f are distributions of each of the six rate terms that make up Inequalities (15) and (17). The signs of the terms are consistent with (15). Figure 4a shows that the rate of change of chemical free energy density in the cells, $\rho^c \dot{\psi}^c_{chem}$, is negative everywhere in the tumor spheroid meaning that the chemical free energy density of the cells is being continually depleted to fuel tumor growth, cell motion, and deformation of the tumor spheroid and gel. Figures 4b and 4c show that the free energy density rates due to cell proliferation, $\pi^c \psi^c$, and due to ECM production, $\pi^e \psi^e$, respectively, are positive in the tumor spheroid since energy is being stored in the newly-formed cells and ECM, and that these terms vanish in the gel where cell proliferation and ECM production are absent. From Figure 4d we see that $-\sigma : d_{gr}$, is positive over the tumor spheroid as cell proliferation and ECM production ensure a positive growth rate (increase in mass with associated swelling) while the stress is compressive (trace of the stress tensor is negative) due to the mechanical constraint of the gel. Therefore,

---

[10] Recall that the cells have no interactions with the agarose hydrogel, which they would if a collagen gel or matrigel were used instead.

*In silico* estimates of free energy changes in growing tumor spheroids the associated change in free energy density is stored, not dissipated as it would be if $-\sigma : d_{gr}$ were negative (equivalently, if $\sigma : d_{gr}$ were positive as discussed in Section 2.3). This rate vanishes over the gel due to the absence of growth there. The rate of change of free energy density due to cell motion, given by $\rho^c \nabla \psi^c \cdot v^c$ in (15), however, is positive, and therefore dissipative as shown in Figure 4e. Its contribution to the rate of change of free energy density is close to zero everywhere except for a rim of motile cells at the tumor-gel interface, where it is negative as the cells move downhill over their own free energy landscape. Figure 4f shows the rate of change of free energy density due to glucose consumption, $\pi^g \psi^g$. The negative values over the tumor spheroid indicate that glucose is being consumed, but the rate decreases sharply to zero over the gel where glucose undergoes diffusion but no consumption.

It also is instructive to compare the magnitudes of the various free energy storing and dissipating mechanisms that have been illustrated in Figure 4. Particularly noteworthy is that the rate of change of the chemical free energy of cells ($\rho^c \dot\psi^c_{chem}$), rate of free energy storage in newly-formed cells and ECM ($\pi^c \psi^c$ and $\pi^e \psi^e$, respectively), and the rate at which free energy is drawn from glucose ($\pi^g \psi^g$) are the same order of magnitude and comparable to each other. This suggests that the (mainly) biochemical processes associated with the cancer's dynamics are relatively close to being balanced as they interconvert free energy. We recall the basis on which these terms were modeled: The chemical free energy rate in the cells, $\dot\psi^c_{chem}$ was adopted from the value noted for metabolic power output of mammalian cells in culture by West et al. (2002) and the power drawn from glucose consumption; see Equation (16). Our models for the source terms $\pi^c$ and $\pi^g$ were abstracted from the work of Casciari et al. (1992) on EMT6/Ro mouse mammary tumor cells, as discussed in Section 2.1, where we also explained our model for $\pi^e$. The elastic moduli, $\kappa$ and $\mu$ that parameterize the Mooney-Rivlin strain energy density function, $W$, for the tumor spheroid were based on estimates for moduli of cells in Suresh (2007) that were extended by assuming elastic incompressibility of the tumor spheroid. The same were assumed for the gel, and were in good correspondence with our initial measurements of the mechanical properties of gels See Section 3.2 and Table 1. The total free energy density of cells was written as

$$\rho^c \psi^c = \rho^c \psi^c_{chem} + \frac{\rho^c}{\rho^c + \rho^e} W, \qquad (18)$$

with West et al.'s (2001) estimate of the chemical energy content of a single mammalian cell being used to specify $\psi^c_{chem}$. For the ECM, however only the strain energy density was included:

$$\rho^e \psi^e = \frac{\rho^e}{\rho^c + \rho^e} W. \qquad (19)$$

The well-known free energy of glucose metabolism was used to specify $\psi^g$ (Table 1).

Also of note is that the rate of free energy density storage due to growth against stress $-\sigma : d_{gr}$ is three orders of magnitude smaller than the terms $\rho^c \dot\psi^c_{chem}, \pi^c \psi^c, \pi^e \psi^e$ and $\pi^g \psi^g$, all of which involve some biochemical processes. This suggests that the mechanical processes that take place in growing tumors are highly energy efficient compared with the biochemical ones. The free energy dissipated during cell motion, $\rho^c \nabla \psi^c \cdot v^c$, is even smaller because the cell diffusivity, $D^c = 10^{-16}$ m$^2$-sec$^{-1}$, which has been *In silico* estimates of free energy changes in growing tumor spheroidsassumed for these cells results in very slow cell motion. Over a day the random motion of motile cells will cause them to displace an average of 7.2 $\mu$m.[11]

*4.1 Parametric studies*
We next considered parametric variations to model biophysically relevant perturbations to the tumor spheroid-gel system. Plots of these results have not been shown here for the want of space, but the essential findings are discussed. The first of these parametric variations is a five-fold increase in cell doubling time, $t_D$ to model the effect of a gene knockout that transforms the cells into a very slowly proliferating phenotype. For this Equation (6) was simply modified by a constant multiplicative factor. The cancer cells proliferate much more slowly to the extent that, at 20 days, $\rho^c$ is approximately half of that in Figure 2. Less ECM is also produced. The more sparsely cellular tumor spheroid consumes correspondingly less glucose.

For this model of a tumor with less proliferative cells we have re-computed the different contributions to the free energy inequality. The rate of change of chemical free energy density in the cells, $\rho^c \dot{\psi}^c_{chem}$, is scaled down by a factor of 3, while the rate of change of free energy density due to ECM production, $\pi^e \psi^e$, and the rate of consumption of the free energy density in glucose, $\pi^g \psi^g$, are scaled down by a factor of approximately 2, influenced mainly by the decreased cell concentration. The decrease in growth rate also causes less free energy storage by the stress power mechanism, $-\sigma : d_{gr}$, attenuating it by a factor of approximately 3. The most strongly affected terms, however, are the rate of storing free energy in newly-formed cells, $\pi^c \psi^c$, and the dissipation of free energy by cell motion, $\rho^c \nabla \psi^c \cdot v^c$. The increased cell doubling time translates to an exponential decrease in $\pi^c$ [see Equation (5)], which with the coupling of $\rho^c$ and $\rho^g$ via the cell proliferation and glucose consumption terms, $\pi^c$ and $\pi^g$, respectively [Equations (5,6,9)], causes an order of magnitude decrease in $\pi^c \psi^c$ and a 20-fold decrease in $\rho^c \nabla \psi^c \cdot v^c$. Notably, the dramatic decrease in free energy dissipated due to cell motion comes about even as the cell flux, $\rho^c v^c$ itself remains unchanged.

Similarly, we have modeled an increase in motility of the tumor cells by increasing the cell diffusivity to $D^c = 10^{-13}$ m$^2$-sec$^{-1}$. The corresponding cell, oxygen and glucose concentrations ($\rho^c, \rho^o, \rho^g$) are shown in Supporting Information as Figures S1—S3. No significant variations are observed in any of the free energy density rate terms except for the dissipation due to cell motion, which increases by three orders of magnitude in direct relation to the magnification of the cell flux, $\rho^c v^c = -D^c \nabla \rho^c$. The larger diffusivity $D^c = 10^{-13}$ m$^2$-sec$^{-1}$ means that the cells displace by ~225 $\mu$m over a day in comparison with a displacement of ~7.2 $\mu$m for $D^c = 10^{-16}$ m$^2$-sec$^{-1}$. This model of higher motility cells is in good quantitative agreement with reports in the experimental literature of highly metastatic glioma cells (Deisboeck et al., 2001; Hegedus et al., 2004). The corresponding diffusivity, $D^c = 10^{-13}$ m$^2$-sec$^{-1}$, also was used for gliomas by Khain and Sander (2006). However, even this greatly magnified dissipation due to enhanced cell motility remains two orders of magnitude lower than the rates of free energy change governed by the dominant biochemical mechanisms discussed above. An increase in gel stiffness by an order of magnitude affected only the rate of free energy storage due to growth against a stress, $-\sigma : d_{gr}$, magnifying it by a factor of nearly 3 while the other rate terms showed no significant variations.

---

[11] At this low value the cell diffusion term in Equation (1) could be neglected. Instead we have relied on the computations to attain the limits of low- and high-motility cells.

*In silico* estimates of free energy changes in growing tumor spheroids

**5. Discussion**

*5.1 Rationale for computational study of free energy changes in growing tumor spheroids*
Firstly, we reiterate our statement that free energy change at the tumor scale is a universal measure for quantification of the physical processes that govern cancer's progression. This has been the motivation for our continuum theory-based mathematical formulation of tumor physics and its extension to obtain a precise statement of the relation between the free energy rates of the various mechanisms at the tumor scale. The free energy inequality (15) is this statement, and it is important to note here that it is a local statement, holding for each point of the tumor spheroid. We note also that it is of interest to have spatially-varying field values of the terms in the free energy inequality. It allows us to study the effect of non-uniform concentrations (of cells, ECM, oxygen and glucose), stress and boundary conditions on the biochemical dynamics and mechanics of the tumor. For instance, it would allow the study of how non-uniform tumor growth and cell motion are induced over the tumor spheroid by these conditions. In addition to the spatial variation in the field values, tumor scale studies would also allow us to investigate how free energy rates change during tumor spheroid development, thereby providing a model for energy usage at different stages of a cancer.

If we were to rely solely on experimental studies to examine these free energy rates for a tumor spheroid system, we would need to obtain spatially-varying values for each of the fields in Inequalities (15) and (17). Experimentally, it is possible to obtain spatially-varying field values for the concentrations, $\rho^c, \rho^e, \rho^o$ and $\rho^g$, and from the changes in $\rho^c$ and $\rho^e$, the rate of deformation tensor due to growth, i.e., $d_{gr}$ can be computed. However, the mass-specific free energies, $\psi^c_{chem}, \psi^c$ and $\psi^e$, must be computed on the basis of constitutive models and the stress, $\sigma$, from the viscoelastic and active stress models for mechanical response. These can be computed very efficiently within the computational model. We note also that the experiments will allow only a limited spatial resolution, i.e., the values can be measured only at a limited number of points. However, with the mathematical model the spatial resolution is limited only by the computational power available, and that provides a resolution that is several orders of magnitude greater than from experiments.[12] Additionally, a mathematical model is necessary for a system-wide understanding of how different (biochemical and mechanical) effects interact and lead to emergent properties. It is for these reasons that we have pursued a computational evaluation of the terms in the free energy inequality. For this specific study of tumor-scale free energy rates, experiments play the critical role of providing parameters that are as precise as possible within the limits of the techniques.

*5.2 The roles of tumor scale and hierarchical studies*
The studies of free energy rates associated with the biochemical and mechanical processes at the tumor scale can be complemented by investigations of the same processes at the cell-ECM and sub-cellular scales. We note that many experimental techniques have been developed for this purpose, while models of cell mechanics and sub-cellular processes are also common. Such studies at a hierarchy of scales would show how the free energy rates of these processes change between the sub-cellular, cell-ECM and tumor scales. This could suggest, for example, whether the agglomeration of single cells into solid tumors causes significant changes in free energy rates associated with cell proliferation or cell motility. Thereby the impact of collective behavior on the development of cancers that form solid tumors can be studied. New insights can thus be gained into the development of the cancer by exploiting the universality of free energy rates as a measure for comparison.

---

[12] We note, of course, that it is meaningless to numerically resolve $\rho^c$ finer than the scale of a single cell (~10 μm), while $\rho^e, \rho^o$ and $\rho^g$ may well be resolved below this scale.

*5.3 Free energy estimates*
The preliminary computations that we have presented here show that the rate of change of the chemical free energy density of cells ($\rho^c \dot{\psi}^c_{chem}$), rate of free energy density storage in newly-formed cells and ECM ($\pi^c \psi^c$ and $\pi^e \psi^e$, respectively), and the rate at which free energy density is drawn from glucose ($\pi^g \psi^g$) are the same order of magnitude, and comparable to each other. This suggests that the rates of free energy interconversion are relatively close to being balanced between the (mainly) biochemical processes associated with the cancer's dynamics. The total rate of change of free energy density is negative as required by (15) and also of the same magnitude as the above terms. The numerical value of this total rate is this preliminary study's estimate of the imbalance in free energy conversions that is being lost in accordance with the second law of thermodynamics. That this loss is comparable to the free energy converted by any of the four dominant tumor scale mechanisms listed above suggests that these biochemical processes that dominate cancer's dynamics are inefficient.

*5.3.1 Relative free energy density rates of biochemical and mechanical deformation processes*
Before considering this interpretation of free energy rates in terms of efficiency it is useful to recall the sources for the parameters that play important roles in this estimate, and reflect on origins of uncertainty or errors therein. The consumption rates of oxygen, $\pi^o$, and glucose, $\pi^g$, are based on models fitted to data for EMT6/Ro cancer cells (Casciari et al., 1992), the optimal cell doubling time, $t_D^{opt}$, was matched to our own preliminary tumor growth experiments with LS174T human colon adenocarcinoma cells, and the rate of change of chemical free energy in cells, $\dot{\psi}^c_{chem}$, was the sum of the estimate for metabolic power output by normal mammalian cells in culture from West et al. (2002), and the energy drawn from glucose consumption. While we have not yet found specific quantitative data for $\dot{\psi}^c_{chem}$ for cancer cell lines, it is quite possible that it may be significantly different from normal cells. Apart from this, the free energy rate terms in (15) must be consistently measured on the same cancer cell line(s) to rigorously validate these findings regarding the mechanisms that dominate the free energy inequality. We emphasize the preliminary, exploratory nature of this finding. In general, more comprehensive experiments are needed from which consistent data can be obtained for a single system of cancer cells. One purpose of our paper is to motivate a comprehensive experimental study of this nature that is forthcoming.

The inefficiency of the biochemical processes suggests that cancer cells do not experience the evolutionary pressure to be energy efficient at the early, pre-vascular stage of the tumor that is represented by tumor spheroids. That the dissipation of a large amount of free energy (inefficiency) is detrimental for survival of cells is reflected in the development of a necrotic core in the tumor spheroid. However, rather than herald a suppression of tumor development, the necrosis only causes termination of the exponential growth phase, as has been well-documented (see the literature cited in this regard in Section 1.2). *In vivo*, the tumor continues to develop via angiogenesis—an adaptation that may be viewed as "subsidizing" the inefficiency of its biochemical processes in the pre-vascular stage.

We have already discussed the origins of the quantities that parameterize the mechanical response of the tumor spheroid, the rates of cell proliferation, oxygen and glucose consumption and cell motion. Additionally, the estimates for moduli of soft matter, including cells, and hydrogels are widely reported (see Suresh, 2007 and references therein; Helmlinger et al., 1997). Given the growth rate of tumor spheroids, also widely reported in the many papers that we have cited, there is a good degree of confidence in the tumor growth rates, $d_{gr}$, and using the moduli, therefore in the stress, $\sigma$. On this basis, the estimates for rate of free energy storage due to growth against stress are accurate at least up to the order of magnitude. We therefore expect that our conclusion is valid on the relative insignificance of the rate at which free energy is stored by this mechanism. Even *in vivo* the stress created in a growing solid

*In silico* estimates of free energy changes in growing tumor spheroids tumor confined by surrounding tissue is unlikely to exceed the stress in our computations by more than an order of magnitude, and we expect that the rate of free energy storage by this mechanism is low *in vivo* also. The small free energy density rate associated with this contribution in comparison with the biochemical processes suggests that there is little prospect of achieving tumor growth control by energy starvation through mechanical interventions. However, such control may be achieved via biochemo-mechanical signaling as suggested by the work of Helmlinger et al. (1997) and Chang et al. (2008). We draw attention to the fact that the tumor growth model does not incorporate the stress-induced suppression of growth that has been observed in these two papers. Inclusion of this effect would only further lower the rate of free energy storage due to growth against stress in the computations.

*5.3.2 Free energy changes associated with cell motion*

The rate of dissipation of free energy due to cell motion, $\rho^c \nabla \psi^c \cdot v^c$, is the scalar product of the cell flux, $\rho^c v^c$, and the gradient of the mass-specific free energy of cells, $\nabla \psi^c$. Of these the magnitude of the cell flux is determined largely by the diffusivity, $D^c$, which we have varied to model nearly immotile to higher motility cells. The mass-specific free energy of cells is $\psi^c = \psi^c_{chem} + W/(\rho^c + \rho^e)$, of which we have estimated $\psi^c_{chem}$ from West et al.'s (2001) calculation of the energy required to form a mammalian cell, $\psi^c_{form}$, and the additional energy gained from glucose consumption:

$$\psi^c_{chem} = \psi^c_{form} + \frac{1}{\rho^c} \int_0^t \psi^g \left(\rho^g(s) - \rho^g_0\right) ds, \qquad (20)$$

where the integral is from the initial time to the current time, and $\rho^g_0$ is the initial glucose concentration. The values of $\psi^c_{chem}$ therefore have some dependence on the solution of the IBVP, and there is a degree of uncertainty arising from the choice of initial and boundary conditions. For this reason there is a little less confidence in our results pointing to the insignificance of the dissipation due to random cell motion. Furthermore, chemotactic and haptotactic cell motion, which are important for eventual metastasis, have not been included in this study. The inclusion of these effects in studies of later (vascularized) stages of tumor development could qualitatively change this estimate.

We note also that the physics of single cell motion is complex. It involves remodeling of the actin cytoskeleton by polymerization and depolymerization, deformation of the cell membrane during filopodial motion, retraction of the cell membrane at the trailing edge due to actomyosin contractility, cell attachment and detachment regulated by focal adhesion dynamics, and mechanical interactions of the cell with the ECM. Free energy changes are involved in all of these phenomena. Some of these are mechanisms of free energy storage in newly-formed actin stress fibers, focal adhesions and their elastically-deforming parts, while free energy is dissipated due to reaction, diffusion and viscous processes. However, none of these effects are represented in our model at the tumor scale where cells are treated as a concentration field. The physics of single cell motion outlined above would emerge in models at the sub-cellular and cell-ECM scales. The dissipation represented in the term $\rho^c \nabla \psi^c \cdot v^c$ only corresponds to the rate of free energy loss as cells migrate to positions in the medium where they possess lower free energy. We will obtain a more complete estimate of the rate of change of free energy density due to cell motion by carrying out studies at a hierarchy of scales: sub-cellular, cell-ECM and the tumor scale. It could reveal whether the mechanisms that govern this process at these different scales introduce notable quantitative and qualitative differences between the rate of change of free energy due to cell motion at the different scales. The present communication only serves to begin this study. Investigations of the free energy resources consumed by cell motion over this hierarchy of scales when carried out for different cell lines and biochemical/mechanical interventions will ultimately lead to a better understanding of the metastatic potential of cancers.

We note also that more strongly adherent cells would have a higher energy of binding with each other and with the ECM. While the free energy changes associated with formation and breakage of adhesions have not been included, we note that higher adhesion energies would manifest themselves in lower cell diffusivity—an effect that we have considered. Also of note in this regard is the work of Turner (2005) where a continuum model was obtained for cell motion by accounting for the energy of cell-cell and cell-ECM adhesion. A multiscale technique was developed, in which the continuum limit of adhesive interactions between discrete cells and the ECM led to a fourth-order nonlinear PDE for transport that bears relation to the Cahn-Hilliard Equation. Khain and Sander (2008) also made direct use of the Cahn-Hilliard Equation to model adhesive effects in cell transport. As explained above, we foresee incorporating the influence of adhesion and other lower scale effects by carrying out studies at a hierarchy of scales to investigate the free energy usage of tumors.

*5.3.3 The role of parametric studies*
The parametric studies that we have carried out by varying the cell doubling time, cell diffusivity and gel stiffness also are of an exploratory nature. Variations in the cell doubling time bring about dramatic changes in the free energy stored in newly-formed cells, $\pi^c \psi^c$, and the free energy dissipated due to cell motion, $\rho^c \nabla \psi^c \cdot v^c$, suggesting that further studies are needed to validate, and if confirmed, to explain the strength of this coupling of biophysical effects. These studies could include both computations such as here but with more comprehensive and consistent data, as well as investigations at the cell-matrix and sub-cellular scales.

Among other avenues for exploration are the parameterization of cell proliferation, metabolic, and oxygen/glucose consumption rates for different gel stiffnesses and cell-ECM adhesiveness. The empirical determination of these rates and their use in the free energy computations that we have demonstrated here will reveal whether and how the mechanisms of free energy change are altered by these different conditions. This also may point to further studies that will have to include the cell-cell, cell-ECM and sub-cellular scales.

*5.4 Onset of necrosis*
The model also can be used to study the onset of necrosis. Figure 5 shows the distributions of $\rho^c$, $\rho^o$ and $\rho^g$ for a more proliferative phenotype, which was modeled by an optimal doubling time $t_D^{opt}$ that is half of the value of the baseline case in Figures 2—4. Note the pronounced depletion of oxygen and glucose, which has been studied in a number of experiments (cited in Section 1.2) for its role in the onset of necrotic cores in tumor spheroids. However, we have not been able to identify a sufficiently precise criterion for onset of necrosis from these experiments. For this reason we have not attempted to model the formation of a necrotic core in our computations, and have concentrated on free energy changes in the pre-necrotic stage. Note that the higher consumption of oxygen and glucose by the more aggressively proliferating cells has also resulted in a noticeable radial gradient of cell concentration: the core being depleted of oxygen and glucose has proliferated slower. This inhomogeneity is made possible by the cell proliferation rate modeled by Equations (5) and (6) with their pointwise dependence on $\rho^o$ and $\rho^g$. These equations were obtained by applying the results of experiments on entire tumor spheroids held at fixed $\rho^o$ and $\rho^g$ (Casciari et al., 1992) to each point in the computational domain. When these equations are solved in our computations, even though each point of the tumor follows an exponential growth law, the distribution of cell proliferation rate is inhomogeneous due to spatial variations in $\rho^o$ and $\rho^g$ (Figure S4 in Supporting Information). While still proliferating with an exponential law, the core is close to the onset of necrosis due to depletion of oxygen and glucose. In future studies we will consider the free energy rates during development of the necrotic core in tumor spheroids—a state which is observed during the development of true solid tumors, also.

*In silico* estimates of free energy changes in growing tumor spheroids

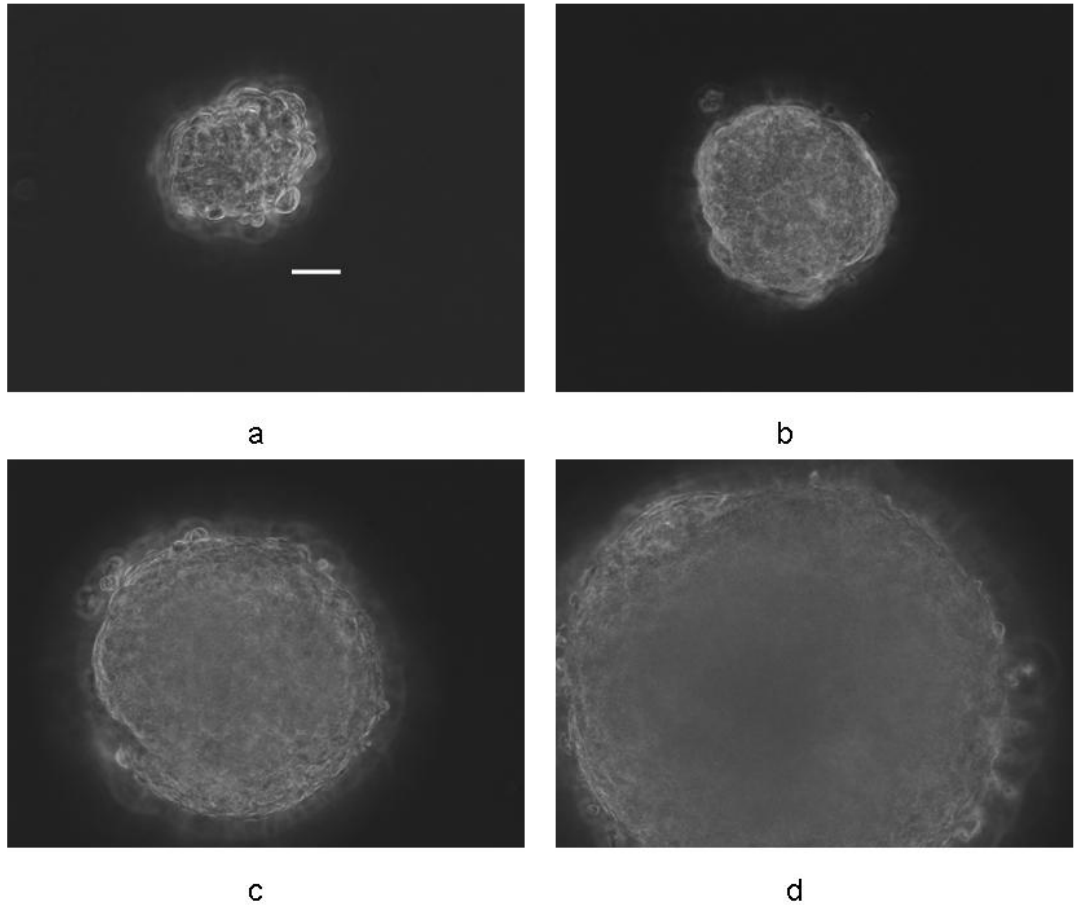

**Figure 1.** Time progression of the growth of an LS174T tumor embedded in a 0.5% agarose gel. Image (a) is a phase image of the spheroid shortly after its transplantation into the gel followed by images taken at (b) 45, (c) 81, and (d) 140 hours after transplantation. The scale bar is 50 μm.

*In silico* estimates of free energy changes in growing tumor spheroids

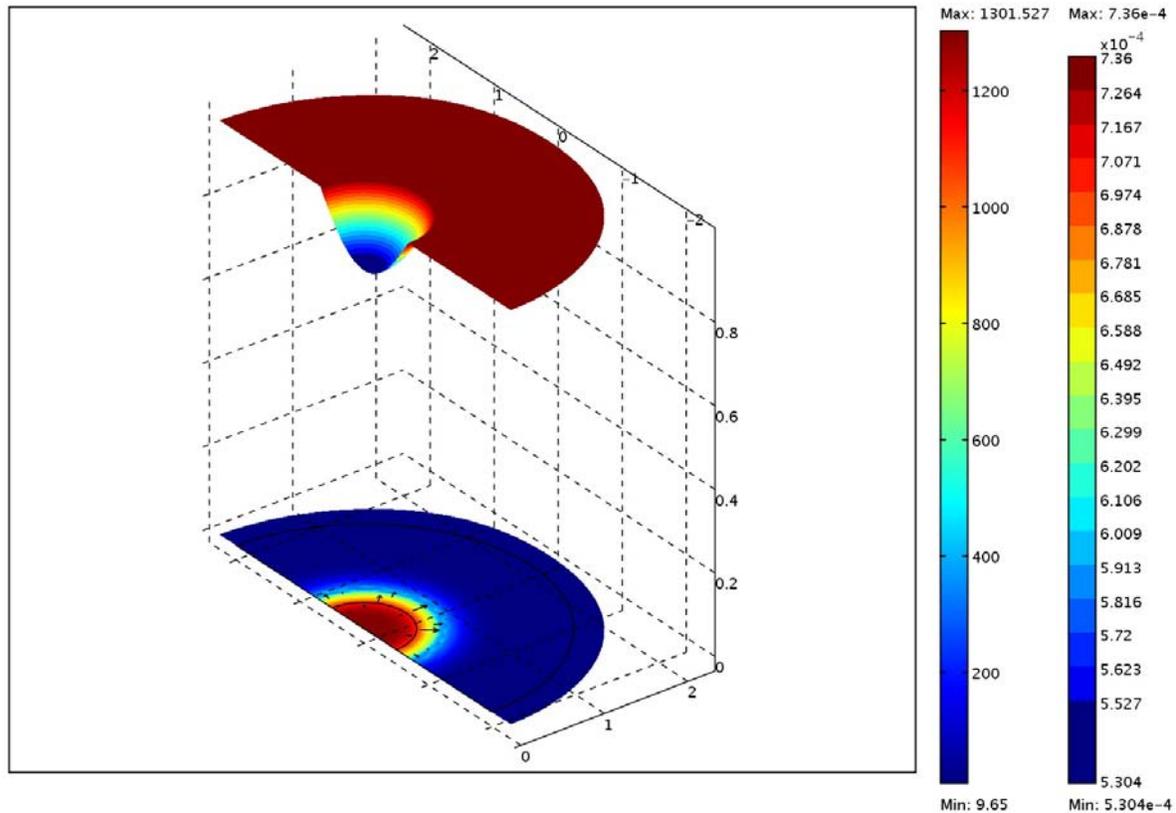

**Figure 2.** Computation of a growing tumor spheroid encapsulated in a gel. The state shown corresponds to 20 days. This two-dimensional model represents a slice through the tumor spheroid and a surrounding gel, and appears as the semicircle in the *lower* part of the figure. The initial configuration of the tumor spheroid is the inner semicircle of 50 μm radius, whose boundary appears as a thin black curve. The ring of initial thickness 150 μm bordering the tumor spheroid is the encapsulating gel against whose mechanical resistance the spheroid grows. The initial extent of the gel is indicated by the outer black semicircle located at 200 μm. The surface plot on the semicircular slice represents cell concentration in mg-cc$^{-1}$ (left hand-side legend). The *upper* surface plot represents oxygen concentration by its colors (right hand-side legend), and glucose concentration by its height, both in mg-cc$^{-1}$. The arrows on the tumor spheroid are flux vectors of the motile cells. The extent of the tumor spheroid is revealed by the high concentration of cells in its core. The cell concentration decreases sharply to zero in the gel.

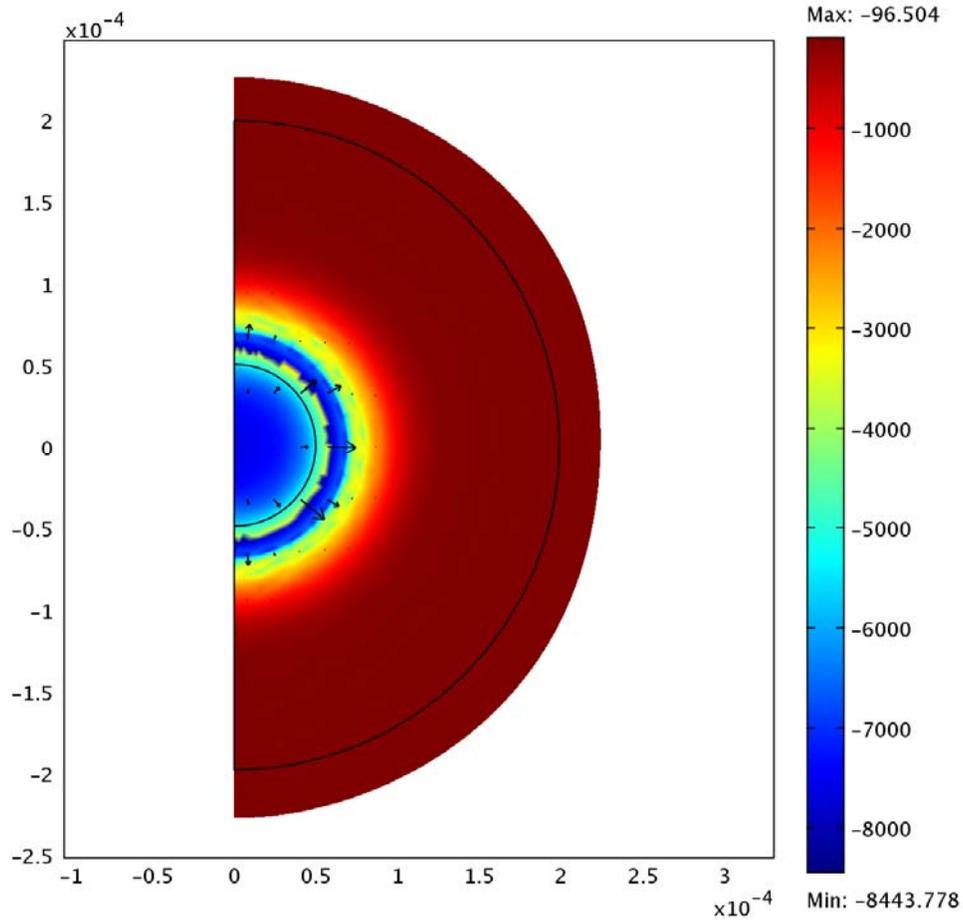

**Figure 3.** The surface plot shows the distribution, over the tumor spheroid and gel, of the rate of change of free energy density from all terms in the free energy inequality (15). The units are Wm$^{-3}$. The arrows show the flux of cell motion.

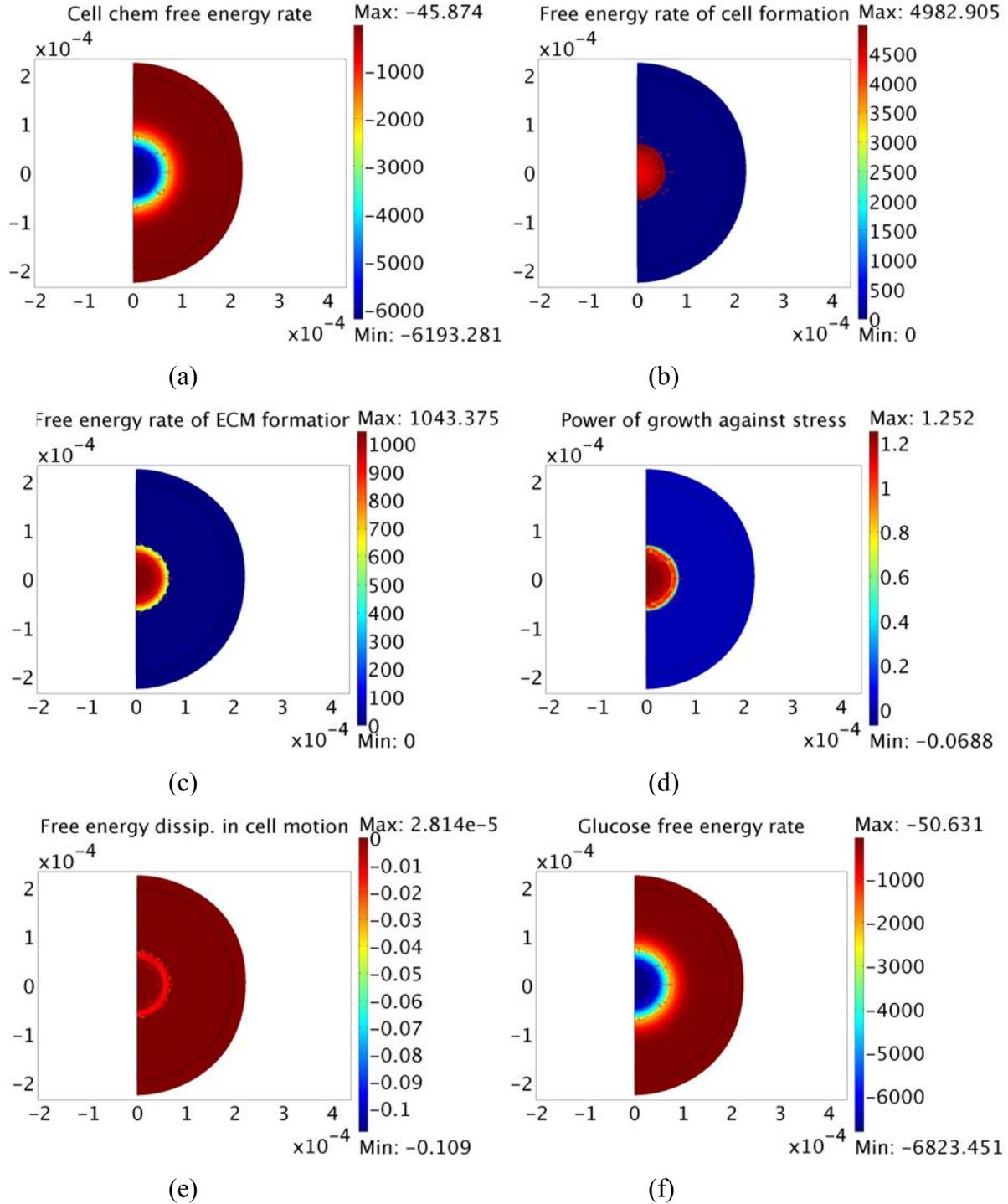

**Figure 4.** Surface plot of the free energy rate terms in (15). All units are Wm$^{-3}$. (a) The rate of change of chemical free energy density stored in the cells, $\rho^c \dot{\psi}^c_{chem}$. (b) The rate of change of free energy density stored in newly formed cells, $\pi^c \psi^c$. (c) The rate of change of free energy density stored in newly-produced ECM, $\pi^e \psi^e$. (d) The rate at which free energy density is dissipated into work done as the tumor spheroid grows against stress, $-\sigma : d_{gr}$.

(e) The rate at which free energy density is dissipated due to cell motion, $\rho^c \nabla \psi^c \cdot v^c$. (f) The rate of change of free energy density due to glucose consumption, $\pi^g \psi^g$.

*In silico* estimates of free energy changes in growing tumor spheroids

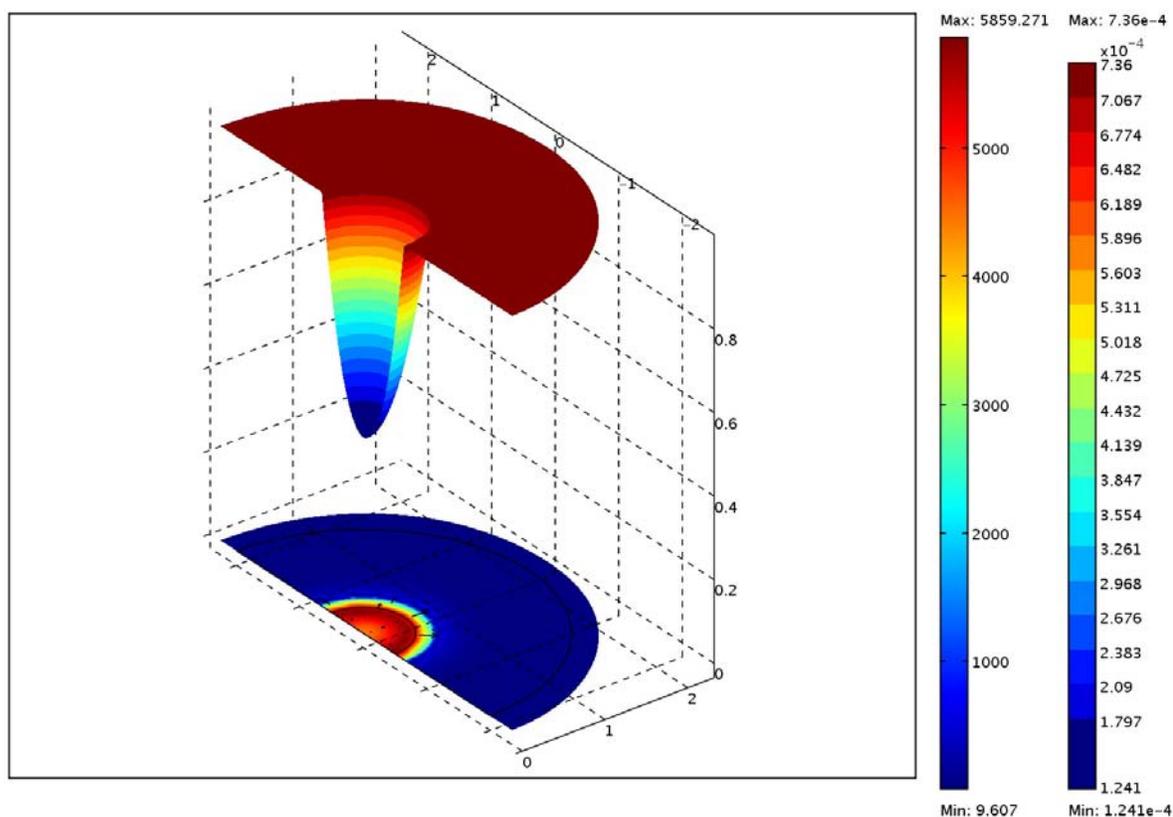

**Figure 5.** The dynamical state of the tumor spheroid at 28.3 days with more proliferative cells. The cell doubling time is half the value of the case modeled in Figure 2. Note the significantly greater depletion of glucose and oxygen in the tumor spheroid's core. This two-dimensional model represents a slice through the tumor spheroid and a surrounding gel, and appears as the semicircle in the *lower* part of the figure. The initial configuration of the tumor spheroid is the inner semicircle of 50 $\mu$m radius, whose boundary appears as a thin black curve. The ring of initial thickness 150 $\mu$m bordering the tumor spheroid is the encapsulating gel against whose mechanical resistance the spheroid grows. The initial extent of the gel is indicated by the outer black semicircle located at 200 $\mu$m. The surface plot on the semicircular slice represents cell concentration in mg-cc$^{-1}$ (left hand-side legend). The *upper* surface plot represents oxygen concentration by its colors (right hand-side legend), and glucose concentration by its height, both in mg-cc$^{-1}$. The arrows on the tumor spheroid are flux vectors of the motile cells. The extent of the tumor spheroid is revealed by the high concentration of cells in its core. The cell concentration decreases sharply to zero in the gel.

**Table 1.** Parameters used in the continuum model.

| Parameter | Units | Value | Remarks |
|---|---|---|---|
| $\kappa$ (tumor spheroid) | Pa | 100000 | Tumor bulk modulus estimated from tumor Young's modulus 6 kPa (Suresh, 2007) in the limit of infinitesimal strain for an incompressible soft material. |
| $\mu$ (tumor spheroid) | Pa | 2013 | Obtained by matching tumor Young's modulus 6 kPa (Suresh, 2007) in the limit of infinitesimal strain for an incompressible soft material. |
| $\kappa$ (gel) | Pa | 100000 | Gel bulk modulus to match tumor spheroid. |
| $\mu$ (gel) | Pa | 2013 | Gel shear modulus = tumor spheroid shear modulus in the linear regime. |
| $\beta$ | Pa(mg-cc$^{-1}$)$^{-2}$ | 0.55 | Scaled up from actomyosin-generated stress of 5.5 kPa on a focal adhesion (Balaban et al., 2001), and accounting for the ratio of ~$10^4$ between cell and total focal adhesion area. |
| $D^c$ | m$^2$-sec$^{-1}$ | $10^{-16}$ | An estimate for low motility cells displacing ~ 3 $\mu$m over a day by random motion. |
| $A$ | (mg-cc$^{-1}$)$^{-1}$sec$^{-1}$ | $8.27 \times 10^{-8}$ | Estimate for a cell producing 5% of its mass in collagen over a week. |
| $D^o$ | m$^2$-sec$^{-1}$ | $16.5 \times 10^{-10}$ | Jiang et al. (2005) |
| $D^g$ | m$^2$-sec$^{-1}$ | $4.22 \times 10^{-11}$ | Jiang et al. (2005) |
| $B_{cell}$ | W | $3 \times 10^{-11}$ | West et al. (2002) |
| $m_{cell}$ | kg | $3 \times 10^{-12}$ | Common estimate for cells |
| $\psi^g$ | J-kg$^{-1}$ | $4.22 \times 10^5$ | Taken to be the free energy change of the glycolysis reaction that converts one molecule of glucose to pyruvate and 2 ATP molecules (Garrett & Grisham, 2005) |

**Table 2.** Apparent modulus values (in kPa) for various Agarose concentrations and strain rates. Note the strain rate effect.

| | | Agarose concentration | | | |
|---|---|---|---|---|---|
| | | 0.5% | 0.75% | 1.0% | 2.0% |
| Strain rate, sec$^{-1}$ | $1 \times 10^{-4}$ | -- | 1.5 | 2.5 | 15 |
| | $5 \times 10^{-4}$ | 0.7 | 1.8 | 3.8 | 22 |
| | $1 \times 10^{-3}$ | 0.8 | 1.9 | 4.0 | 25 |